\pdfoutput=1

\documentclass[11pt]{article}

\usepackage[]{naacl2021}

\usepackage{times}
\usepackage{latexsym}

\usepackage{graphicx}

\usepackage[T1]{fontenc}

\usepackage[utf8]{inputenc}

\usepackage{microtype}

\title{The Flipped Classroom model for teaching Conditional Random Fields in an NLP course}

\author{Manex Agirrezabal \\
  Centre for Language Technology (CST) -
  Department of Nordic Studies and Linguistics\\
  University of Copenhagen / Københavns Universitet\\
  Emil Holms Kanal 2\\
  2300 Copenhagen (Denmark)\\
  \texttt{manex.aguirrezabal@hum.ku.dk}}

\begin{document}
\maketitle
\begin{abstract}
In this article, we show and discuss our experience in applying the flipped classroom method for teaching Conditional Random Fields in a Natural Language Processing course. We present the activities that we developed together with their relationship to a cognitive complexity model (Bloom's taxonomy). After this, we provide our own reflections and expectations of the model itself. Based on the evaluation got from students, it seems that students learn about the topic and also that the method is rewarding for some students. Additionally, we discuss some shortcomings and we propose possible solutions to them. We conclude the paper with some possible future work.
\end{abstract}

\section{Introduction}
In this article we would like to provide experience on developing a flipped classroom environment in a Natural Language Processing course. The most common approach of teaching involves a teacher ``pouring'' knowledge to its students, as if students were plants and knowledge was water.
There is a tendency, though, to make classes more active and interactive, in which communication does not only happen from the teacher to the students, but also from student to student.

Bloom's taxonomy\footnote{\url{https://www.flickr.com/photos/vandycft/29428436431}} \cite{bloom1956taxonomy,anderson2001taxonomy} is a model that describes the cognitive load of different types of work or activities.
On the one hand, some activities, such as remembering or understanding specific characteristics of, for instance, a model of any kind, would be considered to require a low cognitive load. On the other hand, evaluating which model is best by considering the characteristics of each, could be considered to be in a higher level of complexity in Bloom's taxonomy.

If we want our students to be more knowledgeable and reflective, we need to go beyond the lower levels in Bloom's taxonomy. The traditional approach of introducing topics in a lecture would make students remember and understand the covered topics. We believe, though, that with these teaching practices there is a time limitation to go beyond the first two levels of Bloom's taxonomy.


A more efficient approach could be to ask students to work on a set of topics beforehand, get prepared, and then, work on activities that involve a higher cognitive load. These activities could involve applying and analyzing the acquired knowledge to other aspects. In this article, we present our experience in teaching Conditional Random Fields as a Flipped Classroom. We teach this in a course at the Master level about Natural Language Processing (NLP).

The article is structured as follows. First we introduce the flipped classroom method and some challenges. After that, we present the program, course and lecture in which the new teaching method will be used. Then, we discuss the characteristics of the implied student \cite{ulriksen2009implied} and also the evaluation method that we use. Later, we describe the lectures structure and the different activities and we classify them according to Bloom's taxonomy. We then include some discussion, based on the experience. Finally, we conclude the work and suggest some possible future directions.

\section{Flipped Classroom: Are we going to turn around the desks?}

Flipped Classroom \cite{lage2000inverting,brame2013flipping} is a teaching approach in which students get first exposure to the material of a lecture outside of class, and the in-class activities involve applying the learned content. Provision of lecture material can be done in a number of ways, such as reading material, video lectures as slideshows, podcasts, and so on.

Students are expected to do the homework, and the majority of the in-class activities fully depend on that. Because of that, the teacher should make sure that students do their homework, because even though we name it in various ways, watching lectures or reading articles is still homework \cite{nielsen2012five}, and the challenge of making students accomplish with that is still there. A possible idea for making sure that students do the reading homework could be to ask them to do a quiz or reward them somehow.

Apart from challenges regarding students, we may not forget that all the activities, homework, readings, and so on have to be retrieved, selected or produced. Furthermore, as the content covered in class is more complex, the teacher will have to be more prepared. All this results in a larger working load in the preparation of the class and its activities.




\section{Background about program, course and lecture}
In this section, we briefly introduce the M. Sc. program, the course and the specific lecture in which we will be focusing on.

\subsection*{The IT \& Cognition program} 
The IT \& Cognition program at the University of Copenhagen is an international and interdisciplinary
program\footnote{\url{https://studies.ku.dk/masters/it-and-cognition/}} that accepts a small group of students every year. The program covers three main areas: Natural Language Processing, Image Processing and Cognitive Science.

In the first semester, the students acquire the required basic skills for their further development in the specialization area in which they are interested. Scientific Programming, Language Processing I, Cognitive Science I and Vision and Image Processing are mandatory subjects that cover these basic skills.

In the second semester students continue learning about Language Processing and Cognitive Science in further specialized courses, and they also get an introduction to Data Science. Besides, they start their specialization process with an elective course.

The third and forth semesters are devoted to a third course about Cognitive Science and three different electives and after that, students work on their thesis project.

\subsection*{Language Processing 1 and 2 (LP1 and LP2)}
These courses are taught during a whole academic year (two semesters). There is one lecture per week which lasts for two hours. Each course, LP1 and LP2, goes through fourteen weeks in the fall and spring semesters, respectively.

The Intended Learning Outcomes (ILOs) of both courses are quite similar. The main differences are in the degree of depth in which topics are covered. On the one hand, the first course offers basic knowledge about different tasks relevant to Natural Language Processing and their relationship to current society. On the other hand, the second course is more geared towards the development of more advanced algorithms and their application in more specific tasks.

\subsubsection*{Assessment method}
We assess students by asking them to work on a predefined topic. The common procedure is to perform experiments for a relevant NLP task and afterwards, they should write a scientific article reporting on these experiments.

\subsection*{Lecture: Conditional Random Fields}
One of the topics covered in the second Language Processing course covers knowledge about sequence tagging. We cover Hidden Markov Models, Maximum Entropy Markov Models and Conditional Random Fields, as sequential tagging models. This last model will be covered in one and a half sessions. In the first session (2 hours) we will cover theoretical questions and students are expected to understand them. In the following week, there will be one hour, and students will get hands-on practice about Conditional Random Fields.

The lecture itself has the goal of providing the students an understanding of how Maximum Entropy Markov Models (MEMMs) \cite{mccallum2000maximum} and Conditional Random Fields (CRFs) \cite{lafferty2001conditional} make predictions compared to hard classification methods. Additionally, they should understand a common problem of MEMMs (Label Bias problem \cite{bottou-91a,lafferty2001conditional}) and how CRFs solve such limitation. Finally, students should also be able to use CRFs for their own research after the lectures.

\section{The implied student}
The background of our student group \cite{ulriksen2009implied} is very heterogeneous. Some people may have a Computer Science related background, and therefore, sufficient experience in programming. Other students do not have the same background, but they are strong in other aspects, such as linguistics, neuroscience or psychology. Besides, at the time that our analyzed lecture happens, students have already received lectures about programming (first semester), so therefore, this level gap should be significantly smaller.

\section{Evaluation method}



In recent years, the usual teaching practice in the Language Processing series has been lectures. As the goal of this experiment is to check whether the flipped classroom can support students in their learning or not, we will implement this teaching method for the section about Conditional Random Fields, and analyze how students feel about it.

We evaluate this teaching practice by asking students to fill in a survey. In the survey we ask students about their general knowledge about some topics (MEMMs, CRFs, Label Bias problem) but also about whether they would be able to use CRFs for their own work. Please find below the questions that we asked in the questionnaire:

\small
\begin{enumerate}
\setlength\itemsep{-0.2em}
    \item Do you know what a MEMM is (Maximum Entropy Markov Model)?
    \item Do you know what a CRF is (Conditional Random Field)?
    \item Do you know what the Label Bias problem is?
    \item Do you feel capable of using a CRF for your own problems, such as developing a Named Entity Recognition system?
    \item Do you feel that this structure (teaching style) is more rewarding?
    \item Do you feel that this structure is more demanding (mentally)?
\end{enumerate}
\normalsize

The response to these questions could be either ``Yes'', ``Roughly'' or ``No''. Finally, there are two questions related to the specific teaching method, in which we ask students whether the teaching practice is more demanding (mentally) and also whether it is more rewarding, compared to the lecturing approach. The survey was made one week before the lecture day and after the lecture was done.

\section{Activities}
In this section we describe the activities that we made for students before the lectures, during the lectures and after the lectures. The activities will be made public, hoping that they will be useful for other NLP teachers and/or researchers.

\subsection{Before lecture}
Before the lecture, students were asked to watch two video lectures. The two videos were available at the university learning platform and they were made by ourselves. The first video covers Maximum Entropy Markov Models and we introduce them by showing the relationship to Logistic Regression and  Hidden Markov Models. These last models (HMMs) were introduced two weeks before in this same course. We use Maximum Entropy Markov Models as a middle step in order to understand Conditional Random Fields, which are discussed in the second video. We talk about the Label Bias problem and show how this is solved by using global normalization in Conditional Random Fields.

Students should also read the paper that introduces Conditional Random Fields \cite{lafferty2001conditional}\footnote{We believe that including an additional article about Maximum Entropy Markov Models \cite{mccallum2000maximum} is relevant and very helpful for students. Unfortunately, we did not include it this year.}.

\begin{table*}[]
    \centering
\begin{tabular}{c|ccccc}
t     & 0 & 1& 2& 3& 4\\
\hline
WORDS     & My& smartphone& worked& very& well\\
POS tags & PRP& N& VP& MOD& RB\\
Features & \texttt{1,2,0} & \texttt{0,10,0} & \texttt{0,6,1} &\texttt{0,4,0}& \texttt{0,4,0}\\
\end{tabular}
    \label{tab:my_label}
\end{table*}

\subsection{In classroom (online)}
As mentioned before, CRFs will be covered in one and a half lecture sessions. These sessions are held online because of the current situation, following our health authorities requirements.


Before we start the lecture, students are asked whether there are questions regarding the reading and watching activities.
We will briefly recapitulate some aspects, such as where could sequential models like HMM, MEMM or CRFs be useful: Part-of-Speech (POS) tagging, Named Entity Recognition (NER), and besides, any other task that requires the production of tags for a given sequence of elements.

Then, the goal of the remaining time in the session is threefold: (1) to revise and get an understanding of why sequential models such as HMMs, MEMMs or CRFs are more powerful than hard classification models, e.g. Maximum Entropy, Support Vector Machines (SVM), and so on; (2) to make it clear what the Label Bias problem is; and (3) to show how CRFs solve the Label Bias problem by using global normalization. After this session, there will be time to show CRFs working in practice, so that students get hands-on experience.

We describe below four different exercises that students will have to do in small groups (3-4 people). These exercises have an increasing level of complexity, as it will be seen.

\subsubsection*{Exercise 1}
Understand how prediction is made in a Maximum Entropy POS tagger. Students are given one sentence and the POS tags for each word in that sentence. They are told that the model is trained using three very simple features: \texttt{isUpperCase}, \texttt{length\_chars}, \texttt{endsInEd}), and they have to simulate by hand how predictions are made for each word. We also provide a list of 10 possible POS tags. In order to make sure that the concepts about the weight matrix are understood, we ask some checkpoint questions, such as the shape of the weight matrix, provided that the input matrix has a size of \texttt{1x3} and the output matrix has a size of \texttt{1x10}.

Each group of students should provide two outputs to the teacher. Given an input and a weight matrix,\footnote{we also provide the dot product output to save time} they should be able to see which POS tag would be returned by the model. We also ask them to describe, in one sentence with their own words, how the model produces the output for each word.

Considering Bloom's taxonomy \cite{bloom1956taxonomy}, this exercise could be considered an exercise to remember, understand and apply concepts, and thus, in the three lower levels.

\subsubsection{Exercise 2}
In this exercise, students are given the same exact sentence as before. The difference is that the POS tagger with which the students will work is a Maximum Entropy Markov Model (MEMM), and therefore, there is no independent predictions.

Students already learned about the Viterbi algorithm for Hidden Markov Models (HMM). The goal of this exercise is to remember how this algorithm works and also to try to be aware which is the difference between HMMs and MEMMs, i.e. the use of an extended set of features besides single words and tags.

In order to raise that awareness, the exercise is to go through the pseudo-code of the Viterbi algorithm for HMMs \cite[p. 220]{jurafsky2008speech}, understand it and find out which are the specific elements that have to be changed, so that this works with MEMMs. As a hint in order to guess what should be put there, we provide students a trellis of the example above. We highlight one node in that trellis and discuss what its outgoing arcs represent: A probability distribution $P_{S’}(S|0) = [P_1, P_2, ..., P_{k}]$. The value of $k$ depends on the number of states, i.e. output classes.

This exercise requires understanding the Viterbi algorithm for HMMs and understanding what should be modified to make it work with MEMMs. As students have to draw connections between these two models, we consider that the cognitive load, based on Bloom's taxonomy, is higher than in exercise 1.

\begin{figure*}
    \centering
    \includegraphics[width=0.85\textwidth]{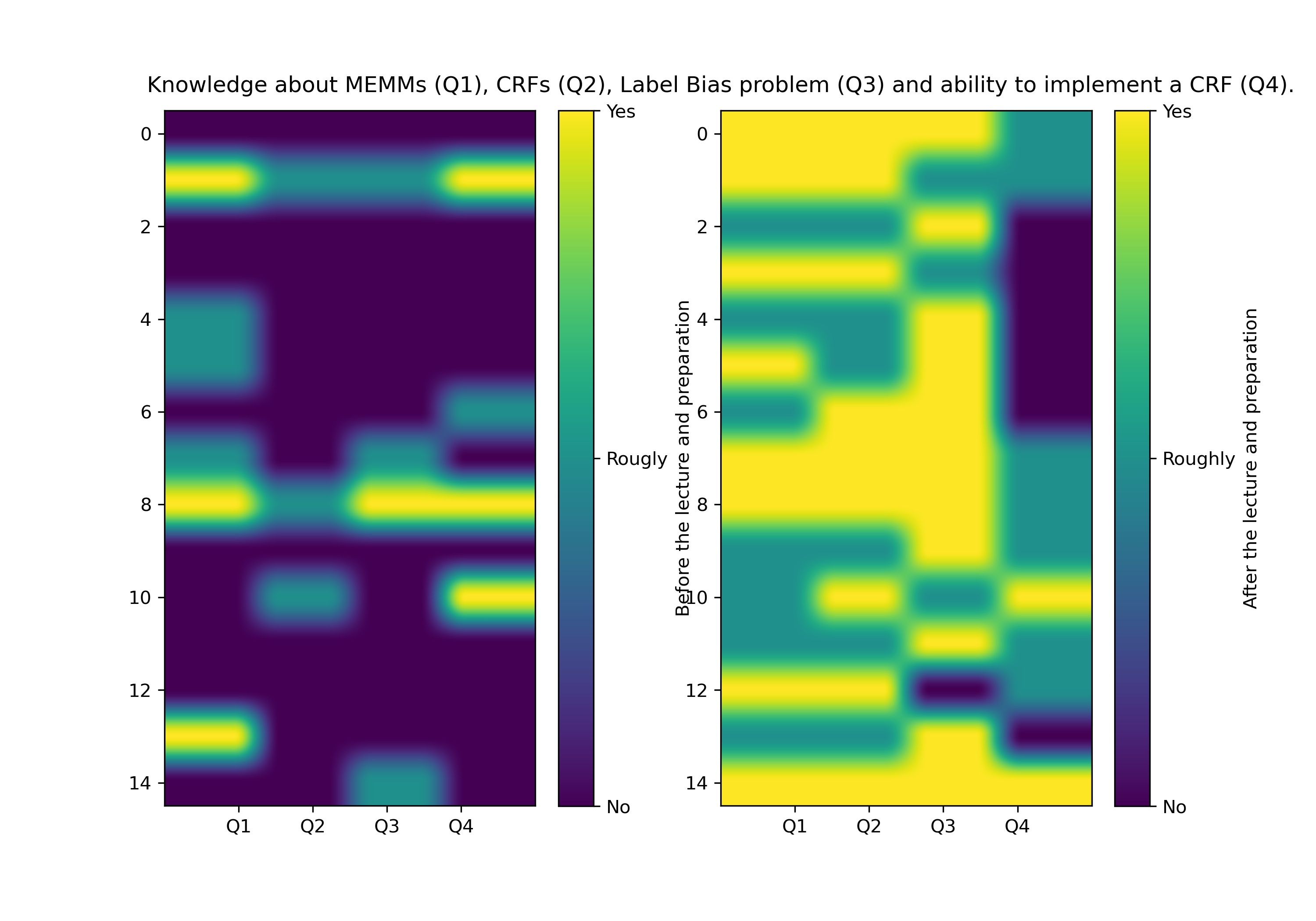}
    \caption{The left plot shows the responses to the survey that we did one week before the flipped classroom session started. The one on the right shows the responses to the same questions, but after the session was over. Based on the responses, we can say that students felt that they understood these four relevant aspects. It looks like, though, that in the future we should emphasize in the practical aspect of the taught models (Look at the right column in the second figure).}
    \label{fig:my_label}
\end{figure*}

\subsubsection*{Exercise 3}
In the previous exercise, when we mention the probability distribution $P_{S'}(S|O)$, we mention that some of those probabilities could be zero. Because of that, in this exercise we ask students to reflect on what would happen if we have a node with many zero probabilities. For example, what happens if a node has 2 non-zero probabilities? What if it has 10 non-zero probabilities? We ask them this question to think about those probabilities and make them aware of the Label Bias problem.

In Bloom's taxonomy this exercise could be seen as evaluating and/or analyzing, as students should be able to find out the problem by themselves. We do not ask them to apply what they know, but to think about how having more or less arcs could affect the probabilities of nodes, and thus, the final sequence probabilities.

\subsubsection*{Exercise 4}
As they already identified the problem, the last exercise is to suggest a solution for that problem. They should analyze the trellis given in exercise 2 and think about a possible solution. Students are given 5-10 minutes to discuss the topic. Afterwards, we discuss their possible solutions and if nobody reaches the actual solution, we introduce global normalization (per observation sequence), which is used in Conditional Random Fields.

In this case, the students would be in a similar scenario as the researchers that found out the Label Bias problem. As such, we consider that this is at the highest level of Bloom's taxonomy, in which students discuss and conjecture about a possible solution for the problem.



\subsubsection*{Practical applications}
After covering the previous exercises, in the next session, we cover Conditional Random Fields from a practical perspective, for which we show students how to process text to use it with Conditional Random Fields. We also discuss how to extract relevant features.

\subsection{After classroom (exam project)}
With regards to the final project, there will be a practical session in which different methods, including CRFs, will be shown. The students will be then able to apply the obtained knowledge.

\section{Discussion and evaluation}
Our expectation is that students will learn more, in a better way with a similar effort (from their side). We think that the reward (for students) of this teaching method is high, on the one hand. On the other hand, the required time for preparation (for teachers) is higher.

The preparation time increase is directly related with the fact that the topics that are covered are expected to be more complex. As students already got lecturing as homework, then the in-class activities become more complicated, and it is because of this that the teacher must have a deeper understanding of the topic in question. Besides, in our specific case the video lectures were made by us. We could save time by using the available resources in online learning platforms, but we decided to make them ourselves, making the preparation more time-consuming.

The evaluation, based on a survey that students had to answer, seems positive, although there are issues. Considering how much students learned, we can say that the learning goals were satisfied, as it can be seen in Figure \ref{fig:my_label}. Each column in the plots represents the answers of our students to a question regarding knowledge on different concepts. They had to answer either Yes, Roughly or No. As we can see, students did not have much expertise on the topic before the lecture and preparation. Afterwards, the responses show that students got the required understanding.

Besides, we also asked whether this teaching method is more rewarding. 7 out of 15 answered yes. There was one that answered no. Finally, there were other 7 out of 15 people that did not take any stance, as they responded “I don’t know”. From these results we cannot strongly confirm that the method, in the way that we implemented it, is rewarding. It seems quite rewarding, though.

Together with these answers, we gave the students the option of writing further comments. There were some positive comments, and some others were possible issues with suggestions for improvements. One mentions that it is hard not to be able to ask questions when you watch the lecture, and that it is hard to remember the context of the question in class. A possible solution to this issue could be to include a discussion forum for each video lecture, so that students could post the question immediately in the forum. Another student pointed out that homework distribution was far from being optimal, as all homework was given in the first week and in the second week there was nothing. This should definitely be thought in a way that the homework load is more balanced. On the bright side, it seemed positive to have the chance of watching again the lecture videos. Also, some felt that group work was better and that it was nice to have more time to understand code and exercises.

\section{Conclusion and Future directions}
In this paper, we presented a possible class structure for teaching Conditional Random Fields in almost two lectures. This class is formulated as a Flipped Classroom, in which there is a strong workload in the students' preparation and this allows students to get a further understanding of the topic, compared to traditional lecturing.

We contemplate the Flipped Classroom as a relevant teaching method for teaching complex topics. We believe that by asking students to do part of the work beforehand has allowed us to go one step beyond in the understanding of CRFs. Furthermore, the exercise types that we covered were in the highest orders in Bloom's taxonomy, showing the efficiency of the method.

The feedback that we got from students seems to show that, in general, they learn about the topic in question. It further seems that the method is rewarding for some students. We believe that the methodology has advantages, e.g. students get a deeper understanding of the topic, and disadvantages, for instance a higher preparation time. Considering both aspects, a possibility could be to find a balance between flipping only a portion of the whole lecture and having the other portion as a more traditional lecture.


In our prepared lectures, we decided to emphasize on a problem that Conditional Random Fields fix, the Label Bias problem. In the future, it would be interesting to include a discussion about the observation bias \cite{klein2002conditional}, which happens when the prediction is made by totally ignoring the labels.

As students have a very varied background, we could already observe differences in the time of execution of the exercises. A possible solution to this is to apply differentiated teaching \cite{rock2008reach}, where the teaching content is adjusted to some groups of students, making the activities more tailored to those student clusters.

\section*{Acknowledgements}
First of all, I would like to acknowledge the whole Teaching and Learning in Higher Education (TLHE) program taught at the University of Copenhagen, especially to my pedagogical and academic supervisors, Lis Lak Risager (TEACH centre) and Patrizia Paggio (Centre for Language Technology), respectively. I would like to thank also the students at Language Processing 2 (Spring semester, course 2020/2021, IT \& Cognition program) for being so helpful in the development of this model. Last, but not least, I thank the anonymous reviewers for their valuable comments and revisions. These reviews are not only useful for this article, but also for the development of the whole course of Language Processing.

\bibliography{custom}
\bibliographystyle{acl_natbib}




\end{document}